\documentclass{cupconf}

  \checkfont{eurm10}
  \iffontfound
    \IfFileExists{upmath.sty}
      {\typeout{^^JFound AMS Euler Roman fonts on the system,
                   using the 'upmath' package.^^J}%
       \usepackage{upmath}}
      {\typeout{^^JFound AMS Euler Roman fonts on the system, but you
                   dont seem to have the}%
       \typeout{'upmath' package installed. cupconf.cls can take advantage
                 of these fonts,^^Jif you use 'upmath' package.^^J}%
      }
  \else
  \fi

  \checkfont{msam10}
  \iffontfound
    \IfFileExists{amssymb.sty}
      {\typeout{^^JFound AMS Symbol fonts on the system, using the
                'amssymb' package.^^J}%
       \usepackage{amssymb}%
         
       \let\ge=\geqslant  
      }{}
  \fi

  \IfFileExists{amsbsy.sty}
    {\typeout{^^JFound the 'amsbsy' package on the system, using it.^^J}%
     \usepackage{amsbsy}}
    {}

\newsavebox{\astrutbox}
\sbox{\astrutbox}{\rule[-5pt]{0pt}{20pt}}

\newcommand\etal{\mbox{\textit{et al.}}}

\newcommand\be{\begin{equation}}
\newcommand{\ee}{\end{equation}}

\title[Black holes, entropy and information]{Black Holes, Entropy, and Information}

\author[G. T. Horowitz]{Gary T. Horowitz}

\affiliation{Physics Department, UCSB, Santa Barbara, CA 93106}

\begin{document}

\maketitle

\begin{abstract}
Black holes are a continuing source of mystery. Although their classical properties have been understood since the 1970's, their quantum properties raise some of the deepest questions in theoretical physics. Some of these questions have recently been answered using string theory. I will review these fundamental questions, and the aspects of string theory needed to answer them. I will then explain the recent developments and new insights into black holes that they provide. Some remaining puzzles are mentioned in the conclusion.
\end{abstract}

\firstsection % if your document starts with a section,
\section{Introduction}

General properties of black holes were studied extensively in the early 1970's, and the basic theory was developed. One of the key results was Hawking's proof that the area of a black hole cannot decrease [\cite[Hawking (1971)]{Hawking:1971tu}]. This led \cite[Bekenstein (1973)]{Bekenstein:1973ur} to suggest that a black hole should have an entropy proportional to its horizon area. This suggestion of a connection between black holes and thermodynamics was strengthened by the formulation of the laws of black hole mechanics [\cite[Bardeen \etal\ (1973)]{Bardeen:1973gs}]. In addition to the total mass $M$, angular momentum $J$, and horizon area $A$ of the black holes, these laws are formulated in terms of the angular velocity of the horizon $\Omega$,  and its surface gravity $\kappa$. Recall that the surface gravity is the force at infinity required to hold a unit mass stationary near the horizon of a black hole. Of course,  the force near the horizon diverges, but there is a redshifting effect so that the force at infinity remains finite. The laws of black hole mechanics are the following:
\vskip .2cm

0) For stationary black holes, the surface gravity is constant on the horizon

1) Under a small perturbation: 
\be\label{first}
dM = {\kappa \over   8 \pi G} dA + \Omega d J
\ee

2) The area of the event horizon always increases

\vskip .2cm
\noindent The zeroth law is obvious for nonrotating black holes which are spherically symmetric, but it is also true for rotating black holes which are not.
If $\kappa$ is like a temperature, and $A$ is like an entropy, then there is a striking similarity to the ordinary laws of thermodynamics:

\vskip .2cm
0) The temperature of an object in thermal equilibrium is constant

1) Under a small perturbation: 
\be\label{firstthermo}
dE = T dS - P dV
\ee

2) Entropy always increases
\vskip .2cm

\noindent At the time it seemed clear that the analogy between black holes and thermodynamics should not be taken too seriously, since if black holes really had a 
temperature, they would have to radiate, and everyone knew that 
nothing could come out of a black hole. Two years later, everything changed. 

\cite[Hawking (1975)]{Hawking:1974sw} coupled quantum matter fields to a classical black hole, and showed that they emit black body radiation with a temperature 
\be
k T = {\hbar \kappa\over 2\pi}
\ee
 So adding quantum mechanics in this limited way (which was 
all that was known how to do) made the analogy complete.
 Black holes really are thermodynamic objects.
For a solar mass black hole, the temperature is very low ($T \sim 10^{-7}$ K) so it is astrophysically negligible. But $T \sim 1/M$ so if a black hole starts evaporating, it gets hotter and eventually explodes. A black hole would start evaporating if it is a 
small primordial black hole formed in the early universe, or if we wait a very 
long time until the three degree background radiation redshifts to less than 
$10^{-7}$ K.

Hawking's determination of the temperature, together with the first law (\ref{first}), fixed the coefficient in Bekenstein's formula for the black hole entropy:
\be\label{sbh}
S_{BH} = {k\over 4 \hbar G} A
\ee
This  is an enormous amount of entropy. A solar mass black hole has $S_{BH} \sim 10^{77} k$. This is much greater than the entropy of the matter that collapsed to form it: Thermal radiation has the highest entropy of ordinary matter, but a
 ball of thermal radiation has          $M \sim T^4 R^3$,       $S \sim T^3 R^3$. 
When it forms a black hole $R \sim M$, so $T \sim M^{-1/2}$ and hence   $ S \sim M^{3/2}$.    On the other hand,    $S_{BH} \sim M^2$. So  $S_{BH}$ grows much faster with $M$ than the entropy of a ball of thermal radiation of the same size. Since we have suppressed all physical constants, the two entropies are equal only when $M$  is of order  the Planck mass ($10^{-5}$ gms). We will continue to set $c=k=\hbar =1$ in the following.

The  discovery that  black holes are thermodynamic objects raised the following fundamental questions:
\vskip .2cm
(1) What is the origin of black hole entropy? In all other contexts, thermodynamics is just an approximation to a more 
fundamental statistical 
description in which the entropy is the log of the number of
microstates. The large entropy indicates that black holes have an enormous number of 
microstates. What are they?

(2) Does black hole evaporation lose information? Does it violate quantum mechanics? 
Hawking argued for three decades that it did.
\vskip .2cm
To understand Hawking's argument, recall that another classical property of black holes established in the 1970's 
 was the uniqueness theorem [\cite[Robinson (1975)]{Robinson:1975bv}]: {\it The only stationary 
(vacuum) black hole solution is the Kerr solution.} You can form a black hole by 
collapsing all kinds of different matter with different multipole moments.
However, after  it settles down, the black hole is completely described by 
only two parameters $M, J$. Wheeler described this by saying
``black holes have no hair". The spacetime outside the horizon retains no memory of what was thrown into the black hole. Now
the radiation emitted by a black hole is essentially thermal. It cannot depend on the matter inside without violating causality or locality.
When the black hole evaporates, $M$ and $J$ are recovered, but the detailed information about what was thrown in is lost. In the language of quantum theory, pure states    appear to evolve into   mixed states. This would violate unitary evolution and hence one of the basic principles of quantum mechanics. 

Hawking argued that the formation and evaporation of a black hole  is very different from burning a book. This may seem like it is destroying information, but quantum mechanically, it can be described by unitary evolution of one quantum state into another.
In principle, all the information in the book can be recovered from the ashes and emitted radiation. 

\section{String theory}

String theory is a promising candidate for both a quantum theory of gravity and a unified theory of all the known forces and matter. One of the main successes of string theory is that it has been able to provide answers to the two fundamental questions above. 
To understand these answers, one  needs
a few basic facts about string theory. (For more detail, see e.g. \cite[Zwiebach (2004)]{Zwiebach:2004tj}.) The first is that when one quantizes
a string in flat spacetime, there are an infinite tower of massive states.
For every integer $N$ there are states with 
\be\label{stringm}
M^2 \sim N/l_s^2
\ee
 where $l_s$ is
a new length scale in the theory set by the string tension. These states
are highly degenerate, and one can show that
the number of string states at excitation level $N\gg 1$ is $e^{S_s}$ where
\be
S_s \sim \sqrt N
\ee
i.e. the string entropy is proportional to the mass in string units. One can
understand this in terms of a simple model of the string as a random walk
with step size $l_s$. As a result of the string tension, the energy 
in the string after $n$ steps is
proportional to its length: $E\sim n/l_s$.
If one can move in $k$ possible directions at each step, the total number of
configurations is $k^n$, so the entropy for large $n$ is proportional to
$n$, i.e. proportional to the energy.

String interactions are governed by a string coupling
constant $g$ (which is determined by a scalar field called the dilaton).
Newton's constant $G$ is related to $g$ and the string length $l_s$
by $G \sim g^2 l_s^2$ in four spacetime dimensions. It is sometimes convenient
to  use string units where $l_s=1$, and sometimes to use Planck units where $G = l_p^2 =1$.
It is important to distinguish them, especially when $g$ changes.
 Since $g$ is in fact determined by a dynamical
field, one can imagine that it changes as a result of a physical process,
e.g. a wave of dilaton passing by. However, it will often be convenient
to assume the dilaton is constant and treat $g$ as just a parameter in 
the theory. In general, physical properties of a state can change
when $g$ is varied. But we will see that
in some cases,  certain 
properties remain unchanged.

The classical
spacetime metric
is well defined in string theory only when the curvature is less than
the string scale $1/l_s^2$. This follows from the fact that fundamentally,
the metric is
unified with all the other modes of the string. This is easily seen in
perturbation theory where the graviton is just one of the massless excitations
of the string. When the curvature is small compared to $1/l_s^2$,
one can integrate out the massive modes and obtain an effective low
energy 
equation of motion which  takes the form of Einstein's equation with an infinite
number of correction terms consisting of higher powers of the curvature
multiplied by powers of $l_s$.  When curvatures approach the string scale,
this low energy approximation breaks down.

String theory includes supersymmetry. Although this symmetry has not yet been seen in nature, there is hope that it will soon be discovered by the Large Hadron Collider being built at CERN. An important consequence of this new symmetry is the following. Supersymmetric theories have a bound on the mass of all states given by their charge which roughly says       $M \ge  Q$. This is called the BPS bound.
States which saturate this bound are called BPS. They have the special property that the mass does not receive any quantum corrections.

Quantizing a string also leads to a prediction that space has more than three dimensions.  This is because a symmetry of the classical string action is preserved in the quantum theory only in a ten spacetime dimensions. The idea that spacetime may have more than four dimensions was first proposed in the 1920's by Kaluza and Klein. Their motivation was to create a unified theory of the two known forces: gravity and electromagnetism. It turns out that a theory of pure gravity in five dimensions reduces  to gravity plus electromagnetism (plus a scalar field) in four dimensions.
The standard explanation for why we do not see the extra dimensions is that they are curled up into a small ball. However recently, it has been suggested that the extra dimensions might be large, but we do not see them because we are confined to live on a 3+1 dimensional submanifold called a ``brane".

In fact, it was realized about ten 
years ago that  string theory is not just a theory of strings.
There are other extended objects called D-branes. These are generalizations of membranes.
They are nonperturbative objects with mass $M\sim 1/g l_s$. But the gravitational field of a D-brane is proportional to  $GM \sim g l_s$ and hence goes to zero at weak coupling. This means that there is a flat spacetime description of these nonperturbative objects. Indeed, they are simply surfaces on which open strings can end. The strings we have been discussing so far have been  topological circles with no endpoints. The dynamics of D-branes at weak coupling is described by open strings (topological line segments) in which the two endpoints are stuck on certain surfaces. Indeed, the D stands for Dirichlet boundary conditions on the 
ends of the string keeping it on the surface, and the surface itself is the ``brane". All the particles of the standard model (e.g., the quarks, leptons, and gauge bosons) are believed to come from these open strings and are confined to these branes. Only the graviton comes from the closed string and is free to move in the bulk spacetime.

Many types of D-branes exist, of various dimensions, and each carries a charge. If the branes are flat (or, more generally, form an extremal surface) and have no open strings attached, they are BPS states.
Excited D-branes (with open strings added) lose energy when two such strings combine to form a closed string. Since the closed string has no ends, it can leave the brane.

\section{Application to black holes}

We now wish to apply string theory to black holes, and answer the two fundamental questions raised in the introduction. We start with the question:
What is the origin of black hole entropy?

For two decades after Bekenstein and Hawking showed that black holes have an entropy, people tried to answer this question with limited success. The breakthough came in a paper by \cite[Strominger and Vafa (1996)]{Strominger:1996sh}. They considered a charged black hole. 
Charged black holes are not interesting astrophysically, but they are interesting theoretically since they satisfy a bound  just like the BPS bound $M \ge Q$.
Black holes with $M = Q$ are called extremal and have zero Hawking temperature. They are stable, even quantum mechanically.
In string theory, extremal black holes are strong coupling analogs of BPS states. One can now do the following calculation:
Start with an extremal black hole and compute its entropy $S_{BH}$. Imagine reducing the string coupling $g$.  When $g$ is very small, one obtains a weakly coupled system of strings and branes with the same charge.
Strominger and Vafa count the number of BPS states in this system at weak coupling and find
\be
N_{BPS} =e^{S_{BH}}
\ee
This is a microscopic explanation of black hole entropy! 
Unlike previous attempts to explain $S_{BH}$, one counts quantum states of a system in flat spacetime where there is no horizon. One obtains a number which, remarkably, is related to the area of the black hole which forms at strong coupling. 

The idea of decreasing the string coupling should be viewed as a (very useful) thought experiment in string theory. In the real world, $g$ is fixed to some value which is difficult to change. The actual value of the string coupling depends on many details about how string theory is connected to the standard model of particle physics and is not yet known. It is likely to be of order unity.

After the initial breakthrough, the agreement  between black holes and a weakly coupled system of strings and D-branes was  extended in many directions. (For a review, see \cite[Peet (2000)]{Peet:2000hn}.) It was shown that the
entropy agrees for extremal charged black holes with rotation. The
entropy also agrees for near extremal black holes with nonzero Hawking temperature.
Since the 
entropy agrees as a function of energy, it is not surprising that the
radiation from the D-branes has the same temperature as the black hole. What 
was surprising was that the total rate of radiation from black holes
agrees with D-branes. (The analog of Hawking radiation for D-branes is just the process of two open strings combining to form a closed string which leaves the branes.) What was truly remarkable was that the deviations from 
black body spectrum also agree! Neither side is exactly thermal.
On the black hole side, these deviations
arise since the radiation has to propagate through the curved spacetime
outside the black hole. This produces potential barriers which give
rise to frequency-dependent greybody factors. On the D-brane side
there are deviations since the modes come from separate 
left and right moving
sectors on the  D-branes. The calculations of these deviations
could not look more different.
On the black hole side, one solves a wave equation in a black hole 
background. The solutions involve hypergeometric functions. On the
D-brane side, one does a calculation in D-brane perturbation theory.
Remarkably, the answers agree.

More recently,  there has been further progress in counting the microstates of charged black holes. A small black hole in string theory has an entropy which is not exactly given by the Bekenstein-Hawking formula (\ref{sbh}). There are subleading corrections coming from higher curvature terms in the action. \cite[Wald (1993)]{Wald:1993nt}  derived  the form of these corrections to black hole entropy in any theory of gravity. 
Recently, it has been shown that for certain extremal black holes the counting of microstates in string theory reproduces the black hole entropy {\it including these subleading corrections} [\cite[Dabholkar (2006)]{Dabholkar:2006iu}]. The corrections are of order the string scale divided by the Schwarzschild radius to some power. 

What about neutral black holes? \cite[Susskind (1993)]{Susskind:1993ws} suggested that there should be a one-to-one correspondence between ordinary excited string states and black holes.  
Start with a highly excited string with mass (\ref{stringm}) and imagine increasing the string coupling $g$.
Since $G \sim g^2 l_s^2$, two effects take place. First,
the gravitational attraction of one part of the string on the other
causes the string size to  decrease. 
Second, since $G$ increases, the gravitational field produced by the string
becomes stronger and the effective
Schwarzschild radius  $GM$ increases in string units. Clearly, for a 
sufficiently large  value of the
coupling, the string forms a black hole. 

Conversely, suppose  one starts with a black hole and decreases the
string coupling. Then the Schwarzschild radius
shrinks in string units and 
eventually becomes of order  the string scale. At this point
the metric is
no longer well defined near the horizon. Susskind suggested that the 
black hole becomes an excited string state.

When I first heard this, I didn't believe it. The first half of the argument
sounded plausible enough, but the second half seemed to contradict the
well known fact that the string entropy is
proportional to the mass while the black hole entropy is proportional to the
mass squared. It turns out that there is a simple resolution of
this apparent contradiction [\cite[Horowitz and Polchinski (1997)]{Horowitz:1996nw}].
 If one changes the string coupling $g$, the string mass is constant in string units, while the black hole mass is constant in Planck units. Thus
$M_s/M_{BH}$  depends on $g$. We expect the transition to occur when the curvature at the horizon of the black hole reaches the string scale. This implies that the Schwarzschild radius $r_0$ is of order the string scale.
Setting $M_s \sim M_{BH}$ when $r_0 \sim l_s$ we find:
\be
 S_{BH}\sim  r_0 M_{BH}\sim  l_s M_s \sim S_s
\ee
So the entropies agree at this ``correspondence point". This agreement between the string entropy and black hole entropy applies to essentially all black holes, including  higher dimensional Schwarzschild black holes, and charged black holes that are far from extremality. 

This leads to a simple picture for the endpoint of black hole evaporation. In Hawking's picture, the black hole evaporated down 
to the Planck scale where the semiclassical approximations being used broke down. In string theory,
the black hole evaporates until it reaches the string scale, at which point it turns into a highly 
excited string. The excited string continues to radiate
until it becomes an unexcited string, i.e., just another elementary particle. The timescale for black hole evaporation is modified slightly. In the black hole phase, $dM/dt \sim T^2$ and $T \sim 1/M$ so the time to evaporate most of the mass is of order $M^3 $ (in Planck units). When the temperature reaches the string scale,  the black hole turns into a highly excited string. After this transition, the temperature stays at the string scale as the
string radiates.

The above argument shows that strings have enough states to reproduce the entropy of all black holes, but the argument is not precise enough to reproduce the entropy exactly, including the factor of $1/4$.  More recently,  \cite[Emparan and Horowitz (2006)]{Emparan:2006it} showed that one can reproduce the entropy of a class of neutral black holes exactly. These are rotating  black holes in five dimensions which have a translational symmetry around one compact direction (as in Kaluza-Klein theory). If one rewrites the solution as a four dimensional black hole, there are charges associated with the Maxwell field coming from the higher dimensional metric. Using various symmetries of string theory, one can map these charges into D-brane charges and count the   microstates in the same way that was done for BPS black holes.

In fact, a slight extension of this argument yields a precise calculation of the entropy of an extremal Kerr black hole  [\cite[Horowitz and Roberts (2007)]{Horowitz:2007xq}]. This black hole has an entropy which is just given in terms of its angular momentum 
\be
S = 2\pi |J|
\ee
Since $J$ is naturally quantized, this is like the entropy of the extremal charged black holes in which the entropy  is again just a function of the quantized charges. It turns out that one can lift an extremal Kerr black hole to five dimensions and map it into the class of neutral black holes who entropy was counted precisely.

We now return to the second fundamental question raised earlier: Do black holes lose information? For charged, near extremal black holes, the weak coupling limit provides a quantum mechanical  system with the same entropy and radiation.
This was a good indication that black hole evaporation would not violate quantum mechanics. However, the case soon became much stronger.

By studying the black hole entropy calculations, \cite[Maldacena (1998)]{Maldacena:1997re} was led to a remarkable 
conjecture now called the gauge/gravity correspondence:
{\it Under certain boundary conditions, string theory (which includes gravity) is completely equivalent to a (nongravitational) gauge theory Òliving at infinityÓ. }
 At first sight this conjecture seems unbelievable. 
How could an ordinary field theory
describe all of string theory? I don't have time to describe the impressive
body of evidence in favor of this correspondence which has accumulated over
the past decade. The obvious differences between string theory and gauge theory are explained by the fact that our intuition about both theories is largely based on weak coupling analyses. Under the gauge/gravity correspondence, when string theory is weakly coupled, gauge theory is strongly  coupled, and vice versa.

This conjecture provides a ``holographic" description of quantum gravity
in that the fundamental degrees of freedom live on a lower dimensional
space. The idea that quantum gravity might be holographic 
was first suggested by 't Hooft and Susskind motivated by
the fact that black hole entropy is proportional to its horizon area. 

The gauge/gravity correspondence has an immediate consequence: The formation and evaporation of small black holes can be described by ordinary Hamiltonian evolution in the gauge theory. It does not violate quantum mechanics.
After thirty years, \cite[Hawking (2005)]{Hawking:2005kf}  finally conceded this point (although his reasons were not directly related to string theory).

Let me conclude with a few open questions:

(1) Can we count the entropy of Schwarzschild black holes precisely? The recent calculation of the extremal Kerr entropy in terms of microstates gives one hope that this may soon be possible.

(2) How does the information get out of the black hole?  What is wrong with Hawking's original argument?
It appears that we will need some violation of locality. In other words, when one reconstructs the string theory from the gauge theory, physics may not be local on all length scales.

However, perhaps the most important open question is:

(3) What is the origin of spacetime? How is it reconstructed from the gauge theory?
How does a black hole horizon know to adjust itself to have area $A = 4G S_{BH}$?

\end{document}